\documentclass[final,authoryear,3p,times]{elsarticle}
\usepackage{url} %
\usepackage{multicol}
\usepackage{sectsty}
\usepackage{dcolumn}
\usepackage{caption}
\usepackage{subcaption} %

\usepackage{graphicx}
\usepackage{amssymb}
\usepackage{amsmath}

\journal{Journal of Theoretical Biology}

\begin{document}

\begin{frontmatter}

    \title{Statistical mechanics of neocortical interactions \\
    Large-scale EEG influences on molecular processes}

    \author[li]{Lester Ingber\corref{cor1}\fnref{label1}}
    \address[li]{Lester Ingber Research, Ashland, OR} \ead{ingber@alumni.caltech.edu}
    \ead[url]{http://www.ingber.com} \fntext[label1]{Corresponding Author}

    \begin{abstract}
        Calculations further support the premise that large-scale
        synchronous firings of neurons may affect molecular processes.
        The context is scalp electroencephalography (EEG) during
        short-term memory (STM) tasks. The mechanism considered is $
        \mathbf{\Pi} = \mathbf{p} + q \mathbf{A} $ (SI units) coupling,
        where $ \mathbf{p} $ is the momenta of free $ \mathrm{Ca}^{2+} $
        waves $ q $ the charge of $ \mathrm{Ca}^{2+} $ in units of the
        electron charge, and $ \mathbf{A} $ the magnetic vector
        potential of current $ \mathbf{I} $ from neuronal minicolumnar
        firings considered as wires, giving rise to EEG\@. Data has
        processed using multiple graphs to identify sections of data to
        which spline-Laplacian transformations are applied, to fit the
        statistical mechanics of neocortical interactions (SMNI) model
        to EEG data, sensitive to synaptic interactions subject to
        modification by $ \mathrm{Ca}^{2+} $ waves.
    \end{abstract}


\end{frontmatter}

\section{Introduction to top-down premise (TDP)}
This project is motivated by a top-down premise (TDP) that large-scale
synchronous neural firings, as measured by scalp electroencephalographic
recordings (EEG), may directly influence molecular processes that are
coupled to underlying synaptic processes that contribute to this
synchrony
\citep{Ingber2011,Ingber2012,Ingber2015,Ingber+Pappalepore+Stesiak2014,Nunez+Srinivasan+Ingber2013}.

Previous papers
\citep{Ingber2015,Ingber+Pappalepore+Stesiak2014} have primarily been
concerned with fitting EEG data to the author's statistical mechanics of
neocortical interactions (SMNI) model
\citep{Ingber1982,Ingber1983} to study top-down large-scale synchronous
neuronal firings during short-term memory (STM) tasks, also modeled by
SMNI
\citep{Ingber1984,Ingber1985a,Ingber1994}, on coupled bottom-up molecular
processes.

Only for brevity, unless otherwise stated, dependent on the context,
``EEG'' will refer to either the measurement of synchronous firings
large enough to measurable on the scalp, or to the firings themselves.
The term ``molecular processes'' is used to signify not only the scale
included, but also the effects of $ \mathrm{Ca}^{2+} $ waves on synaptic
background molecular processes that ultimately drive neuronal firings
via releases of quanta of chemical transmitters, e.g., affecting
molecules of phospholipid bilayers at presynaptic neuronal membrane
sites. ``Top-down'' refers the largest scale considered of highly
synchronous neuronal firings as measured by EEG, and ``bottom-up''
refers to the smallest scale considered here of $ \mathrm{Ca}^{2+} $
waves as they affect background synaptic processes that contribute to
the neuronal firings measured by EEG in the context of STM tasks.

Free regenerative $ \mathrm{Ca}^{2+} $ waves, arising from
astrocyte-neuron interactions, couple to the magnetic vector potential $
\mathbf{A} $ produced by these collective firings. As calculated in
these papers, only $ \mathbf{A} $, not electric $ \mathbf{E} $ or
magnetic $ \mathbf{B} $ fields have logarithmic insensitivity from their
sources to add cumulatively to affect the molecular processes.
Calculations described here, using data processsed for proper matching
to the SMNI model, give more support to TDP.

This study sheds light on the multiple scales of neocortical
interactions underlying Consciousness ($ \mathbf{C} $). It is now
accepted by some neuroscientists and confirmed by some experiments
\citep{Asher2012,Salazar+Dotson+Bressler+Gray2012}, that at least some
memories are actively processed by highly synchronized patterns of
neuronal firings as measured by scalp EEG recordings during activity of
processing such patterns, e.g., P300 waves, etc. This was always the
basic premise of the SMNI model.

Without detailed experimental testable facts on the nature of $ \mathbf{C}
$, many people consider TDP as just conjecture on other future theories
of $ \mathbf{C} $. As discussed in a previous paper
\citep{Ingber2015} there are aspects of $ \mathbf{C} $ that we may only
be able to infer existence or possibly prove we cannot know. These
latter possibilities can be considered as belonging to a ``Dark $
\mathbf{C} $'' ($ \mathbf{DC} $) category, and $ \mathbf{DC} $ should be
researched as well as $ \mathbf{C} $.

There are many other people that consider $ \mathbf{C} $ as an emergent
property of a complex system, and in this context TDP is a strong
indicator of multiple scales of neocortical interactions underlying $
\mathbf{C} $.

There are many papers that deal directly with or give credence to
advances in $ \mathbf{C} $ that include quantum mechanics, that have
passed a sometimes volatile peer-review process in this subject in
various physics journals
\citep{,Zhou+Mowrey+Tang+Xu2015}, which has led to some critical reviews
\citep{McKemmish+Reimers+McKenzie+Mark+Hush2009}. A current summary gives
fair credit to the new ideas presented
\citep{Atmanspacher2015}.

However, this study demonstrates how models of $ \mathbf{C} $ can be
developed faithful to experimental data. The scientific focus on
computational models that include experimental data opens these ideas to
testable hypotheses.

Results of SMNI fits to EEG data gave strong confirmation of the SMNI
model of STM, and now also give weak statistical support to a basic
physical mechanism that couples highly synchronous firings to control
underlying molecular processes, the canonical momentum $ \mathbf{\Pi } $,
$ \mathbf{\Pi }=\mathbf{p}+q\mathbf{A} $, where $ \mathbf{p} $ is the
momenta of a $ \mathrm{Ca}^{2+} $ wave, $ q $ the charge of $ \mathrm{Ca}^
{2+} $, $ q=-2e $, $ e $ the magnitude of the charge of an electron.

Previous papers have used classical physics to calculate and compare the
molecular $ \mathbf{p} $ and large-scale q $ \mathbf{A} $ components of $
\mathbf{\Pi } $, demonstrating that indeed they of comparable magnitudes
\citep{Ingber2011,Ingber2012,Ingber+Pappalepore+Stesiak2014,Nunez+Srinivasan+Ingber2013}.
Also, in the context of quantum mechanics, the wave function of the $
\mathbf{\Pi } $ system was calculated, and it was demonstrated that
overlap with multiple collisions during the observed long durations of
typical $ \mathrm{Ca}^{2+} $ waves
\citep{Ingber2015,Ingber+Pappalepore+Stesiak2014} support a Zeno effect
\citep{Facchi+Lidar+Pascazio2004,Facchi+Pascazio2008,Giacosa+Pagliara2014,Kozlowski+Caballero-Benitez+Mekhov2015,Patil+Chakram+Vengalattore2015,Wu+Wang+Yi2012,Zhang+Ai+Li+Xu+Sun2014}
promoting long coherence times. This approach also suggests some
nanosystem-pharmaceutical applications
\citep{Ingber2015}.

Note that this proposed quantum context of the $ \mathbf{\Pi } $
interaction does not depend on any quantum effects via internal energy
levels, for which $ \mathrm{Ca}^{2+} $ with atomic weight 40 likely
would not support at room temperatures
\citep{Beck2008}.

Much of the theoretical development in this paper concerns calculations
that support the influence of EEG on $ \mathrm{Ca}^{2+} $ waves, and
referencing experimental support of how $ \mathrm{Ca}^{2+} $ waves
affect background synaptic noise. However, the actual fits to EEG data,
comprising several CPU-years of calculations on parallel-processor
supercomputers, concerns the importance of the background synaptic noise
that includes functional dependence on widespread regional synchrony
arising from logarithmic insensitivity of $ \mathbf{A} $ to distance
from minicolumnar currents. This functional dependence is in the sole
argument of the probability distribution of quanta release which
directly influences molecular processes underlying this EEG\@.
\section{Theory}
\subsection{Molecular processes contributing to synaptic interactions}
There are many studies on tripartite neuron-astrocyte interactions
\citep{Pereira+Furlan2009}, and on $ \mathrm{Ca}^{2+} $ waves at
tripartite sites. The short summary below is presented to set the
context for SMNI calculations of probability distributions of synaptic
activity that include background contributions.

Several studies have shown that glutamate release from astrocytes
through a $ \mathrm{Ca}^{2+} $-dependent mechanism can activate
receptors located at the presynaptic terminals. Regenerative
intercellular calcium waves (ICW) can travel over 100's of astrocytes,
encompassing many neuronal synapses. These ICWs are documented in the
control of synaptic activity
\citep{Scemes+Giaume2006}. Analysis of fluorescence accumulation clearly
demonstrate that glutamate is released in a regenerative manner, with
subsequent cells that are involved in the calcium wave releasing
additional glutamate
\citep{Innocenti+Parpura+Haydon2000}.

Several experiments, albeit with sometimes contradicting conclusions,
support astrocyte generated $ [\mathrm{Ca}^{2+}] $ ($ [\ldots ] $ =
concentration) influences on synapses
\citep{Volterra+Liaudet+Savtchouk2014}. They conclude that new
experimental paradigms are required to address external interferences,
e.g., possible interference of measurement indicators, etc., wherein $ [\mathrm
{Ca}^{2+}] $ affect increased release probabilities at synaptic sites,
likely due to triggering release of gliotransmitters. It has been noted
by other investigators that intracellular astrocyte calcium waves in
situ (not the free $ \mathrm{Ca}^{2+} $ waves considered here),
increases neuronal firings
\citep{Fiacco+McCarthy2004}, and that $ \mathrm{Ca}^{2+} $ waves in
general may have various influences on neuronal firings
\citep{Agulhon+Petravicz+McMullen+Sweger+Minton+Taves+Casper+Fiacco+McCarthy2008}.
It also has been suggested that by simultaneously activating and
deactivating neurotransmission in all of the synapses enveloped by an
astrocyte, the astrocyte calcium wave may coordinate synapses into
synchronously firing groups
\citep{Mitterauer+Kofler-Westergren2011}.

Although the full set of mechanisms affecting $ [\mathrm{Ca}^{2+}] $ and
the influences of $ \mathrm{Ca}^{2+} $ on other mechanism are not yet
fully understood and experimentally verified, it is clear that $ \mathrm
{Ca}^{2+} $ waves exist in intercellular as well as in intracellular
media
\citep{Ross2012}. There are regenerative as well as non-regenerative
processes observed, both ``locally'' at cellular sites as well as into
``expanded'' regions through which $ \mathrm{Ca}^{2+} $ travel at
relatively fast velocities for large distances over relatively long
periods of time
\citep{Ingber2015,Ingber+Pappalepore+Stesiak2014}. $ \mathrm{Ca}^{2+} $
affects spontaneous synaptic production of glutamate, in contrast to
also possibly influencing evoked production due to neuronal firings.
\subsection{Statistics of synaptic interactions}
The probability of a neuron firing, in the context of the previous
Section, is based on the statistics of quantal releases of chemical
neurotransmissions across synaptic gaps. This probability can be
determined by a folding of two distributions, a distribution of quantal
release of neurotranmitters across a given synapse, $ \Psi $, and a
distribution over all synapses and the properties of neuron that affect
its firing or not firing, $ \Gamma $ (discussed in the next Section).

As calculated previously
\citep{Ingber1982,Ingber1983}, the interaction of neuron $ k $ ($ k=1,N^{*}
$) with neuron $ j $ across all $ jk $ synapses is defined by a
distribution $ \Psi $ describing $ q $ chemical quanta with mean
efficacy

\begin{equation}
    a_{jk}^{*}=\frac{1}{2}A_{jk}^{*}(\sigma _{k}+1)+B_{jk}^{*}\approx0.01
\end{equation}
with $ B_{jk}^{*} $ a background spontaneous contribution, where $
\sigma _{k}=1 $ if $ k $ fires; $ \sigma _{k}=-1 $ if $ k $ does not
fire. Synaptic efficacy is a measure of ionic permeability, and an
inverse measure of electrical impedance. As detailed previously
\citep{Ingber1982,Ingber1983}, efficacies $ A_{jk}^{*} $ measure chemical
synaptic activity, while efficacies $ B_{jk}^{*} $ measure background
influences. It was shown that the final results of folding $ \Gamma $
and $ \Psi $ distributions is independent of choosing $ \Psi $ to be
Gaussian or Poisson
\citep{Ingber1982,Ingber1983}, generalizing earlier work
\citep{Shaw+Vasudevan1974}.

For example, if $ \Psi $ is Poisson,

\begin{equation}
    \Psi =\exp (-a_{jk}^{*})(a_{jk}^{*})^{q}/q!
\end{equation}
This $ \Psi $ is essentially the probability discussed in papers
referenced in the previous Section, due to triggering release of
gliotransmitters, which may be influenced by $ \mathrm{Ca}^{2+} $ waves
arising from neuron-astrocyte tripartite interactions. $ B_{jk}^{*} $ is
the variable that will be considered a function of $ \mathbf{A} $ below,
i.e., the TDP mechanism detailing how top-down activity measured by EEG
during STM tasks may influence molecular $ \mathrm{Ca}^{2+} $ waves.
E.g., $ \mathrm{Ca}^{2+} $ waves can control glutumate production which
can control the distribution of $ q $ quanta released at synaptic gaps.
This is consistent with $ \mathrm{Ca}^{2+} $ waves from astrocytes as
contributing a diffuse control of neuronal firings via $ B_{jk}^{*} $,
rather than the more direct control of $ A_{jk}^{*} $ due to presynaptic
firings.
\subsection{Scale to neuronal firing}
A Gaussian distribution $ \Gamma $ describes the average intra-neuronal
distribution of electrical polarization across the various neurons in a
minicolumn,

\begin{equation*}
    \Gamma =(2\pi q\phi _{jk}^{2})^{-1/2}\exp [-(W_{jk}-qv_{jk})^{2}]/(2q\phi
    _{jk}^{2}),
\end{equation*}

\begin{equation}
    \lim \limits_{q\rightarrow 0}\Gamma \equiv \delta (W_{jk}),
\end{equation}
where parameters in $ \Gamma $ are specified below.

Using the probability of developing potential $ W_{jk} $ from $ k $,

\begin{equation}
    S_{jk}=\,\sum \limits_{q=0}\limits^{\infty }\,\Gamma \Psi
\end{equation}
and the probability $ S_{j} $ of developing $ W_{jk} $ from all afferent
neurons

\begin{equation}
    S_{j}=\,\int \,\ldots \,\int \,dW_{j1}\ldots dW_{jN^{*}}S_{j1}\ldots
    S_{jN^{*}}\delta \left (W_{j}-\,\sum \limits_{k}\,W_{jk}\right )
\end{equation}
the derived probability for neuron $ j $ to fire, given its interaction
with $ k=1,\ldots ,N^{*} $ neurons is
\citep{Ingber1982,Ingber1983}

\begin{equation*}
    p_{\sigma _{j}}=\,\int \limits_{V_{j}}\limits^{\infty }\,dW_{j}S_{j}\simeq\exp
    (-\sigma _{j}F_{j})/[\exp F_{j}+\exp (-F_{j})]
\end{equation*}

\begin{equation}
    F_{j}=\frac{V_{j}-\,\sum \limits_{k}\,a_{jk}^{*}v_{jk}}{\left ((\pi
    /2)\,\sum \limits_{k^\prime{}}\,a_{jk^\prime{}}^{*}(v_{jk^\prime{}}^
    {2}+\phi _{jk^\prime{}}^{2})\right )^{1/2}}
\end{equation}
Note the dependence of $ F_{j} $ on synaptic parameters which will be
discussed again below.
\subsection{Scale to mesocolumnar neuronal interactions}
Mesocolumns are defined as the dynamics of divergence from minicolumnar
processes, and convergence to regional and macrocolumnar processes.
Statistical mechanics of neocortical interactions (SMNI) was developed
to explicitly model these dynamics
\citep{Ingber1982,Ingber1983}. The math requires a nonlinear stochastic
calculus, first formulated within the context of gravity by many authors
\citep{Cheng1972}, and subsequently generalized to other classical
physics systems by many other authors
\citep{Langouche+Roekaerts+Tirapegui1982}. At this stage SMNI is a
zero-fit-parameter theory, in that all parameters are picked from within
experimentally determined ranges. However, three basic models were then
developed with slight adjustments of the parameters
\citep{Ingber1984}, changing the firing component of the
columnar-averaged efficacies $ A_{jk} $ within experimental ranges as
discussed below.

Using this theory as a guide, discoveries were made that indeed modeled
various aspects of neocortical interactions, e.g., properties of STM $
-- $ e.g., capacity (auditory $ 7\pm 2 $ and visual $ 4\pm 2 $),
duration, stability , primacy versus recency rule, Hick's law,
nearest-neighbor minicolumnar interactions within macrocolumns
calculating rotation of images, etc
\citep{Ingber1982,Ingber1983,Ingber1984,Ingber1985a,Ingber1994}. SMNI was
also scaled to include mesocolumns across neocortical regions to fit EEG
data, as it used here as well
\citep{Ingber1997a,Ingber2012}.

The resulting mathematics is used here for SMNI modeling of EEG data,
further generalized to include possible interactions with $ \mathrm{Ca}^
{2+} $ molecular processes. Using $ G= \{ E,I \} $ to represent
independent excitatory $ E $ and inhibitory $ I $ processes, in the
prepoint (Ito) representation the SMNI Lagrangian $ L $ is

\begin{equation*}
    L=\sum \limits_{G,G^\prime{}}(2N)^{-1}(\dot{M}^{G}-g^{G})g_{GG^\prime
    {}}(\dot{M}^{G^\prime{}}-g^{G^\prime{}})/(2N\tau )-V^\prime{}
\end{equation*}

\begin{equation*}
    g^{G}=-\tau ^{-1}(M^{G}+N^{G}\tanh F^{G})
\end{equation*}

\begin{equation*}
    g^{GG^\prime{}}=(g_{GG^\prime{}})^{-1}=\delta _{G}^{G^\prime{}}\tau ^
    {-1}N^{G}\mathrm{sech}^{2}F^{G}
\end{equation*}

\begin{equation}
    g=\det (g_{GG^\prime{}})
\end{equation}
where $ N^{G} $ = \{$ N^{E}=160 $, $ N^{I}=60 $\} was chosen for visual
neocortex, \{$ N^{E}=80 $, $ N^{I}=30 $\} was chosen for all other
neocortical regions, $ M^{G^\prime{}} $ and $ N^{G^\prime{}} $ in $ F^{G}
$ are afferent macrocolumnar firings scaled to efferent minicolumnar
firings by $ N/N*\approx 10^{-3} $, and $ N* $ is the number of neurons
in a macrocolumn, about $ 10^{5} $. $ \tau $ is usually considered to be
on the order of 5-10~msec; this is further discussed below in the
Section below on coarse-graining EEG data.

Moving averages of several epochs of $ g^{G} $ are used as slower drifts
as described below in the data Section.

The threshold factor $ F^{G} $ is derived as

\begin{equation*}
    F^{G}=\sum \limits_{G^\prime{}}\frac{\nu ^{G}+\nu ^{\ddagger E^\prime
    {}}}{\big((\pi /2)[(v_{G^\prime{}}^{G})^{2}+(\phi _{G^\prime{}}^{G})^
    {2}](\delta ^{G}+\delta ^{\ddagger E^\prime{}})\big)^{1/2}}
\end{equation*}

\begin{equation*}
    \nu ^{G}=V^{G}-a_{G^\prime{}}^{G}v_{G^\prime{}}^{G}N^{G^\prime{}}-\frac
    {1}{2}A_{G^\prime{}}^{G}v_{G^\prime{}}^{G}M^{G^\prime{}}
\end{equation*}

\begin{equation*}
    \nu ^{\ddagger E^\prime{}}=-a_{E^\prime{}}^{\ddagger E}v_{E^\prime{}}^
    {E}N^{\ddagger E^\prime{}}-\frac{1}{2}A_{E^\prime{}}^{\ddagger E}v_{E^\prime
    {}}^{E}M^{\ddagger E^\prime{}}
\end{equation*}

\begin{equation*}
    \delta ^{G}=a_{G^\prime{}}^{G}N^{G^\prime{}}+\frac{1}{2}A_{G^\prime{}}^
    {G}M^{G^\prime{}}
\end{equation*}

\begin{equation*}
    \delta ^{\ddagger E^\prime{}}=a_{E^\prime{}}^{\ddagger E}N^{\ddagger
    E^\prime{}}+\frac{1}{2}A_{E^\prime{}}^{\ddagger E}M^{\ddagger E^\prime
    {}}
\end{equation*}

\begin{equation}
    a_{G^\prime{}}^{G}=\frac{1}{2}A_{G^\prime{}}^{G}+B_{G^\prime{}}^{G}\:,\:a_
    {E^\prime{}}^{\ddagger E}=\frac{1}{2}A_{E^\prime{}}^{\ddagger E}+B_{E^\prime
    {}}^{\ddagger E}
\end{equation}
where $ \{ A_{G^\prime{}}^{G},B_{G^\prime{}}^{G},A_{E^\prime{}}^{\ddagger
E},B_{E^\prime{}}^{\ddagger E} \} $, $ A_{G^\prime{}}^{G} $ is the
columnar-averaged direct synaptic efficacy, $ B_{G^\prime{}}^{G} $ is
the columnar-averaged background-noise contribution to synaptic
efficacy. $ A_{G^\prime{}}^{G} $ and $ B_{G^\prime{}}^{G} $ have been
scaled by $ N*/N\approx 10^{3} $ to keep $ F^{G} $ invariant. Other
values taken are consistent with experimental data, e.g., $ V^{G}=10 $~mV,
$ v_{G^\prime{}}^{G}=0.1 $~mV, $ \phi _{G^\prime{}}^{G}=0.03^{1/2} $~mV.
The ``$ ^{\ddagger } $'' parameters arise from regional interactions
across many macrocolumns.

Note that the mesoscopic threshold funcion $ F^{G} $ has similar
functional dependencies on neuronal parameters as the individual
threshold function $ F_{j} $. This provides an audit trail back to
neuronal parameters when $ F^{G} $ is used in further scaling to
regional dynamics to fit scalp EEG data.
\subsubsection{Centering Mechanism (CM)}
As mentioned above, three basic models were developed with slight
adjustments of the parameters
\citep{Ingber1984}, changing the firing component of the
columnar-averaged efficacies $ A_{G^\prime{}}^{G} $ within experimental
ranges, which modify $ F^{G} $ threshold factors to yield (a) case EC,
dominant excitation subsequent firings in the conditional probability,
or (b) case IC, inhibitory subsequent firings, or (c) case BC, balanced
between EC and IC. Furthermore, a Centering Mechanism (CM) on case BC
yields case $ \mathrm{BC}^\prime{} $ wherein the numerator of $ F^{G} $
only has terms proportional to $ M^{E^\prime{}} $, $ M^{I^\prime{}} $
and $ M^{\ddagger E^\prime{}} $, i.e., zeroing other constant terms by
resetting the background parameters $ B_{G^\prime{}}^{G} $, still within
experimental ranges. This has the net effect of bringing in a maximum
number of minima into the physical firing $ M^{G} $-space. The minima of
the numerator then defines a major parabolic trough,

\begin{equation}
    A_{E}^{E}M^{E}-A_{I}^{E}M^{I}=0
\end{equation}
about which other SMNI nonlinearities bring in multiple minima
calculated to be consistent with STM phenomena. In this recent project
\citep{Ingber2015,Ingber+Pappalepore+Stesiak2014}, a Dynamic CM (DCM)
model is used as well, wherein the $ B_{G^\prime{}}^{G} $ are reset
every few epochs of $ \tau $.
\subsection{Scale to regions}
Large EEG databases have been used to test scaled SMNI at relatively
large regional scales
\citep{Ingber1997a,Ingber1998}. as well as in this project
\citep{Ingber2015,Ingber+Pappalepore+Stesiak2014}. The above Lagrangian $
L $ is used across regions, interacting via myelinated fibers
represented by $ M^{\ddagger E^\prime{}} $ firings.

The context for SMNI describing EEG was developed using the
Euler-Lagrange (EL) equations derived from the variational principle
associated with SMNI Lagrangians at different scales, giving rise to
basic dynamic variables

\begin{equation*}
    \mathrm{Mass}=g_{GG^\prime{}}=\frac{\partial ^{2}L}{\partial (\partial
    M^{G}/\partial t)\partial (\partial M^{G^\prime{}}/\partial t)}
\end{equation*}

\begin{equation*}
    \mathrm{Momentum}=\Pi ^{G}=\frac{\partial L}{\partial (\partial M^{G}/\partial
    t)},
\end{equation*}

\begin{equation*}
    \mathrm{Force}=\frac{\partial L}{\partial M^{G}}
\end{equation*}

\begin{equation}
    \mathrm{F-ma}=0:\:\delta L=0=\frac{\partial L}{\partial M^{G}}-\frac
    {\partial }{\partial t}\frac{\partial L}{\partial (\partial M^{G}/\partial
    t)}
\end{equation}
The momenta $ \Pi ^{G} $ define Canonical Momenta Indicators (CMI) which
were used to advantage in previous SMNI papers fitting EEG
\citep{Ingber1997a,Ingber1998,Ingber+Pappalepore+Stesiak2014}, proving to
give superior graphs for analysis to those generated from raw electric
potentials $ \Phi $. The EL equations are identified with $ \mathrm{F-ma}=0
$. This was summarized in a recent paper
\citep{Nunez+Srinivasan+Ingber2013}, detailing EL equations at three
scales:
\subsubsection{Columnar EL}
Macrocolumnar nearest-neighbor minicolumnar interactions were calculated
to include spatial diffusion terms in the Lagrangian defining the
conditional probability distribution of mesocolumnar firings. The EL
equations were derived from this distribution. Linearization of the
space-time EL equations permit the development of stability analyses and
dispersion relations in frequency-wave-number space
\citep{Ingber1982,Ingber1983,Ingber1985b}, leading to wave propagation
velocities of interactions over several minicolumns, consistent with
experiments. This calculation first linearizes the EL, then takes
Fourier transforms in space and time variables. A calculation supports
observed rotation of images in STM\@. The earliest studies simply used a
driving force $ J_{G}M^{G} $ in the Lagrangian to model long-ranged
interactions among fibers
\citep{Ingber1982,Ingber1983}.
\subsubsection{Strings EL}
The Lagrangian defining the conditional probability distribution of
mesocolumnar firings was transformed to a conditional probability
distribution of changes in measured EEG electric potentials. The EL
equations were derived from this distribution. This calculation
considered one firing variable along the parabolic trough, discussed
above, of attractor states being proportional to $ \Phi $, the EEG
electric potential
\citep{Ingber+Nunez1990}. There exist regions in neocortical parameter
space such that the nonlinear string model often used to model EEG is
recovered as this EL equation. In this recent study reported here,
spline-Laplacian transformations on the $ \Phi $ are considered
proportional to the firing variables at each electrode site.
\subsubsection{Springs EL}
Macrocolumnar nearest-neighbor minicolumnar interactions were calculated
to include spatial diffusion terms in the Lagrangian defining the
conditional probability distribution of mesocolumnar firings. The EL
equations were derived from this distribution. Some SMNI studies
included in calculations regional interactions driving localized
columnar activity within these regions
\citep{Ingber1997a,Ingber1998}, instead of the crude model using a
driving force $ J_{G}M^{G} $ in the Lagrangian as described above. This
extension of the above EL equations describes EEG oscillatory behavior
supported at these columnar scales across regions
\citep{Ingber2009,Ingber+Nunez2010}, which is a model of coupled
oscillatory springs across macrocolumns and regions, supported within
the string-model envelope described above.
\subsection{Influence of $ q\mathbf{A} $ on $ \mathbf{p} $}
\subsubsection{Classical physics of $ \mathbf{\Pi } $}
Previous papers have modeled minicolumns as wires which support neuronal
firings, due largely from large neocortical excitatory pyramidal cells
in layer V (of six), giving rise to currents which give rise to electric
potentials measured as scalp EEG
\citep{Ingber2011,Ingber2012,Nunez+Srinivasan+Ingber2013}. This gives
rise to a magnetic vector potential

\begin{equation}
    \mathbf{A}=\frac{\mu }{4\pi }\mathbf{I}\log \left (\frac{r}{r_{0}}\right
    )
\end{equation}
which has an insensitive log dependence on distance. In the brain, $ \mu
\approx\mu _{0} $, where $ \mu _{0} $ is the magnetic permeability in
vacuum $ =4\pi 10^{-7} $ H/m (Henry/meter), where Henry has units of
kg-m-C$ ^{-2} $, which is the conversion factor from electrical to
mechanical variables. For oscillatory waves, the magnetic field $
\mathbf{B=\nabla \times A} $ and the electric field $ \mathbf{E}=\frac{ic}
{\omega }\mathbf{\nabla \times \nabla \times A} $ do not have this log
dependence on distance. The magnitude of the current is taken from
experimental data on dipole moments $ \mathbf{Q}=|\mathbf{I}|z $ where $
\mathbf{\hat{z}} $ is the direction of the current $ \mathbf{I} $ with
the dipole spread over $ z $. $ \mathbf{Q} $ ranges from 1~pA-m = $ 10^{-12}
$~A-m for a pyramidal neuron
\citep{Murakami+Okada2006}, to $ 10^{-9} $~A-m for larger neocortical
mass
\citep{Nunez+Srinivasan2006}. These currents give rise to $ q\mathbf{A} $
on the order of $ 10^{-28} $~kg-m/s. $ \mathbf{p} $ from one $ \mathrm{Ca}^
{2+} $ ion in a wave is typically on the order of $ 10^{-30} $~kg-m/s,
and this can be multiplied by the number of ions in a wave, e.g., 100's
to 1000's.
\subsubsection{Quantum physics of $ \mathbf{\Pi } $}
Previous papers also have detailed quantum calculations of the wave
function of $ \mathrm{Ca}^{2+} $ waves in the presence of $ \mathbf{A} $
\citep{Ingber2015,Ingber+Pappalepore+Stesiak2014}. The wave function in
coordinate space, $ \psi (\mathbf{r},t) $ is

\begin{equation*}
    \psi (\mathbf{r},t)=(2\pi  \hbar )^{-3/2}\int \limits_{-\infty }\limits^
    {\infty }d^{3}\mathbf{p}\phi (\mathbf{p},t)e^{i\mathbf{p} \cdot
    \mathbf{r}/ \hbar }
\end{equation*}

\begin{equation*}
    \psi (\mathbf{r},t)=\alpha ^{-1}e^{-\beta /\gamma -\delta }
\end{equation*}

\begin{equation*}
    \alpha =(2 \hbar )^{3/2}(2\pi (\Delta \mathbf{p})^{2})^{3/4}\left (\frac
    {it}{2m \hbar }-\frac{1}{4(\Delta \mathbf{p})^{2}}\right )^{3/2}
\end{equation*}

\begin{equation*}
    \beta =\left (\mathbf{r}-\frac{q\mathbf{A}t}{m}-\frac{i \hbar \mathbf{p}_
    {0}}{2(\Delta \mathbf{p})^{2}}\right )^{2}
\end{equation*}

\begin{equation*}
    \gamma =4\left (\frac{it \hbar }{2m}+\frac{\hbar ^{2}}{4(\Delta \mathbf
    {p})^{2}}\right )
\end{equation*}

\begin{equation}
    \delta =\frac{\mathbf{p}_{0}^{2}}{4(\Delta \mathbf{p})^{2}}+\frac{iq^
    {2}\mathbf{A}^{2}t}{2m \hbar }
\end{equation}
where $ (\Delta \mathbf{p})^{2} $ is the variance of $ \mathbf{p} $ in
the wave packet, and various properties were calculated and shown to be
reasonable in this neocortical context.

If we consider the above wave packet in momentum space, $ \phi (\mathbf{p},t)
$ being ``kicked'' from $ \mathbf{p} $ to $ \mathbf{p}+\mathbf{\delta p}
$, and simply assume that random repeated kicks of $ \delta \mathbf{p} $
result in $ <\delta \mathbf{p}>\approx 0 $, and each kick keeps the
variance $ \Delta (\mathbf{p}+\delta \mathbf{p})^{2}\approx \Delta (\mathbf
{p})^{2} $, then the overlap integral at the moment $ t $ of a typical
kick between the new and old state is

\begin{equation*}
    <\phi ^{*}(\mathbf{p}+\mathbf{\delta p},t)|\phi (\mathbf{p},t)>=e^{\frac
    {i\kappa +\rho }{\sigma }}
\end{equation*}

\begin{equation*}
    \kappa =8\mathbf{\delta p}\Delta \mathbf{p}^{2} \hbar m(q\mathbf{A}+\mathbf
    {p}_{0})t-4(\mathbf{\delta p}\Delta \mathbf{p}^{2}t)^{2}
\end{equation*}

\begin{equation*}
    \rho =-(\mathbf{\delta p} \hbar m)^{2}
\end{equation*}

\begin{equation}
    \sigma =8(\Delta \mathbf{p} \hbar m)^{2}
\end{equation}
where $ \phi (\mathbf{p}+\mathbf{\delta p},t) $ is the normalized wave
function in $ \mathbf{p}+\mathbf{\delta p} $ momentum space. A crude
estimate is obtained of the survival time $ A(t) $ and survival
probability $ p(t) $
\citep{Facchi+Pascazio2008},

\begin{equation*}
    A(t)=<\phi ^{*}(\mathbf{p}+\mathbf{\delta p},t)|\phi (\mathbf{p},t)>
\end{equation*}

\begin{equation}
    p(t)=|A(t)|^{2}
\end{equation}
These numbers yield:

\begin{equation}
    <\phi ^{*}(\mathbf{p}+\mathbf{\delta p},t)|\phi (\mathbf{p},t)>=e^{i
    (1.67\times 10^{-1}t-1.15\times 10^{-2}t^{2})-1.25\times 10^{-7}}
\end{equation}
Even many repeated kicks do not appreciably affect the real part of $
\phi $, and these projections do not appreciably destroy the original
wave packet, giving a survival probability per kick as $ p(t)\approx
\exp (-2.5\times 10^{-7})\approx 1-2.5\times 10^{-7} $. Both
time-dependent phase terms in the exponent are sensitive to time scales
on the order of 1/10 sec, scales prominent in STM and in synchronous
neural firings measured by EEG\@. This suggests that $ \mathbf{A} $
effects on $ \mathrm{Ca}^{2+} $ wave functions may maximize their
influence on STM at frequencies consistent with synchronous EEG during
STM by some mechanisms not yet determined.
\subsubsection{PATHTREE}
A sub-project under the current XSEDE.org grant is developing a
complex-number version of PATHTREE, an algorithm developed by the author
for path integration of financial options
\citep{Ingber+Chen+Mondescu+Muzzall+Renedo2001}, also being developed to
run on parallel processors under OpenMP. PATHTREE can be used to develop
the wave-function above, adding ``shocks'' to the wave packet to
investigate the duration of the wave-packet due to a Zeno/bang-bang
effect. This is similar to the use of PATHTREE to include dividends on
the underlying asset in financial options. PATHINT is another code
developed by the author for path integration used in several
disciplines, including the SMNI project
\citep{Ingber+Nunez1995}, but PATHTREE is much faster than PATHINT.
\section{Data}
\subsection{Previous data}
Previous papers in this project used EEG data given to the author circa
1997
\citep{Ingber1997b,Zhang+Begleiter+Porjesz1997,Zhang+Begleiter+Porjesz+Litke1997,Zhang+Begleiter+Porjesz+Wang+Litke1995}.
This data was used in other projects
\citep{Ingber1997a,Ingber1998}, as well as in previous calculations in
this project
\citep{Ingber2015,Ingber+Pappalepore+Stesiak2014}.

The first use of the data in this project led to examination of multiple
graphs to determine differences between no-$ \mathbf{A} $ and $ \mathbf{A}
$ models
\citep{Ingber+Pappalepore+Stesiak2014}. This was inconclusive, but
leaning to better results with the $ \mathbf{A} $ model. A second
attempt using this data
\citep{Ingber2015} used statistical measures on the cost functions using
the Lagrangian $ L $
\begin{quote}
    \{STAT\} = \{mean, standard-deviation, skewness, kurtosis\}
\end{quote}
applied to each model among the three paradigms presented to each
subject, according to whether the subject was classified as \{a =
alcoholic, c = control (non-alcoholic)\}, and according to paradigm \{1
= single stimulus, m = attempt to match second stimulus to first, n = no
second stimulus matched first\}. This also was inconclusive, but leaning
to better results with the $ \mathbf{A} $ model. Further examination
showed that several trials had cost functions much larger than the rest,
which reasonably could be considered severe outliers skewing results.
Rather than ``cherry-picking'' runs in the data, the decision was made
to look for more recent data.
\subsection{Choice of data}
New data was sought that would satisfy the conditions of the SMNI model,
e.g., scalp EEG during STM task, among a reasonable number of subjects,
each with a reasonable number or runs that could be divided into
Training and Testing sets. A dataset was found with 245 runs across 12
subjects with thousands of epochs per run
\citep{Citi+Poli+Cinel2010,Goldberger+Amaral+Glass+Hausdorff+Ivanov+Mark+Mietus+Moody+Peng+Stanley2000}.
which was downloaded from the
\begin{quote}
    http://physionet.nlm.nih.gov/pn4/erpbci
\end{quote}
site, which also contains the useful link
\begin{quote}
    http://www.biosemi.com/download/Cap\_coords\_all.xls
\end{quote}
that was used for spline-Laplacian transformations described below.

The data used was collected during P300 event-related potentials (ERP),
in the context of tasks designed to measure changes in shape of the ERP
across attentional tasks to non-targets and single and multiple targets
via a Donchin speller algorithm. This speller chooses targets from a
matrix of 36 characters that the subject must input. Multiple
presentations aid in reducing noise of the measured ERP. The authors
give rigorous proof and calculations to support the superiority of their
Donchin speller algorithm over previous experimental setups
\citep{Citi+Poli+Cinel2010}.
\subsection{Conversion to text files}
The data is in European Data Format (EDF) format which contains a lot of
information not directly used in this project where simple ascii data is
used by fast C codes to process long optimization sessions. A very
useful code was found at the site
\begin{quote}
    http://www.teuniz.net/edf2ascii/
\end{quote}
\subsection{Spline-Laplacian transformation}
Arguments have been made, based on a wave-equation analysis of EEG, that
Laplacian transformed data are better than the original raw scalp
electric potential to represent localized source currents giving rise to
EEG
\citep{Srinivasan+Winter+Nunez2006}.

Ramesh Srinivasan gave me part of his Matlab code on
\begin{quote}
    http://ssltool.sourceforge.net
\end{quote}
for spline-Laplacian scalp transformations. We then worked together to
get this section of code to run under Octave, available from
\begin{quote}
    http://www.gnu.org/software/octave/
\end{quote}
using the xls file referenced above which contains coordinates of the
scalp EEG electrode sites. This smaller code is available at no charge
from this author, but any use should reference the ssltool site above.

A typical two-dimensional Laplacian is the weighted difference between
four nearest-neighbor points and a central point. Spline fits are
smooth-fitting algorithms to a set of discrete points. Here, a spline
first fits all (64) sites on the (semi-)circular scalp, generating
piecewise smooth curves, so there is actually input from all points
before the Laplacian is taken. Smoothing data points before taking
derivatives, e.g., expanding the number of points used to calculate the
Laplacian
\citep{Lynch1992}, is a good counter-measure to introducing noise
whenever derivative operations are used on noisy data, as would occur if
numerical Laplacians were used. In this case the spline fits develops an
analytic function with smooth Laplacians. The Laplacian permits a better
localization of sources, since there is a general diffusion of electric
potential especially due to the scalp and skull. There is dependence on
the spline fits to accomplish a lot but still remaining faithful to the
data, e.g., properly representing distant regions of cortex included at
each region-site, but studies support this approach as better than
others
\citep{Srinivasan+Winter+Nunez2006}.

After conversion to text files, each run had a data file of about 70,000
lines, each representing about 0.5 msec. Using Octave, the
spline-Laplacian transformation was applied across 64 columns
representing electrode sites for each of these lines of data across all
245 runs.

When applied to P300 data here, the spline-Laplacian transformed data
were similar, but more tightly grouped, than the original scalp electric
potential data.
\subsection{Coarse-graining time resolution}
The SMNI model is bast on time resolutions refractory period of $ \tau
\approx5 $~msec is taken to lie between an absolute refractory period of
$ \approx1 $~msec, during which another action potential cannot be
initiated, and a relative refractory period of $ \approx0.5--10 $~msec.
Therefore this data was then coarse-grained with a moving average of 4
epochs, with about 17,500 lines representing time resolutions of about 2
msec. The use of a narrow moving average to reduce noise was used
effectively in copula risk management of financial markets, as discussed
in a generalized approach to applications of SMNI
\citep{Ingber2007}, as originally reported in the finance literature
\citep{Litterman+Winkelmann1998}. This method worked better in these
contexts than more sophisticated algorithms developed using random
matrices
\citep{Laloux+Cizeau+Bouchaud+Potters1999}.
\subsection{Cumulative graphs select time window}
All 245 runs across all epochs were placed on one graph. Consistent with
the experimental design, this showed several regions of cumulative high
amplitudes during which STM tasks were performed by the 12 subjects. A
region of continuous high amplitude was chosen of 2561 lines
representing times from 17 to 22 secs.
\subsection{Cumulative graphs select electrode sites}
For each of 64 electrode sites, cumulative graphs over all 245 runs for
all subjects \{s01 ...  s12\} were examined to select sites with obvious
strong signals. The four sites \{05-F3, 37-AFz, 40-F4, 48-Cz\} showed
similar strong signals with two main peaks and one main valley. Other
sites had varying degrees of much larger spreads of scattered data.
Figure 1 is an example comparison from electrode site 05-F3, using
cumulative data over 20 subjects, each with 20+ runs, displaying a graph
of the electric potential data measured at one electrode across all
subjects across all runs, versus a graph of the moving-averaged (discussed
above) spline-Laplacian data. The reduction of noise is due to the
moving average and spline algorithms, and the general shape change is
due to the Laplacian algorithm applied to the entire set of 64
electrodes.

\begin{figure}
    \begin{subfigure}[b]
        {0.4\textwidth} \includegraphics[width=\textwidth]{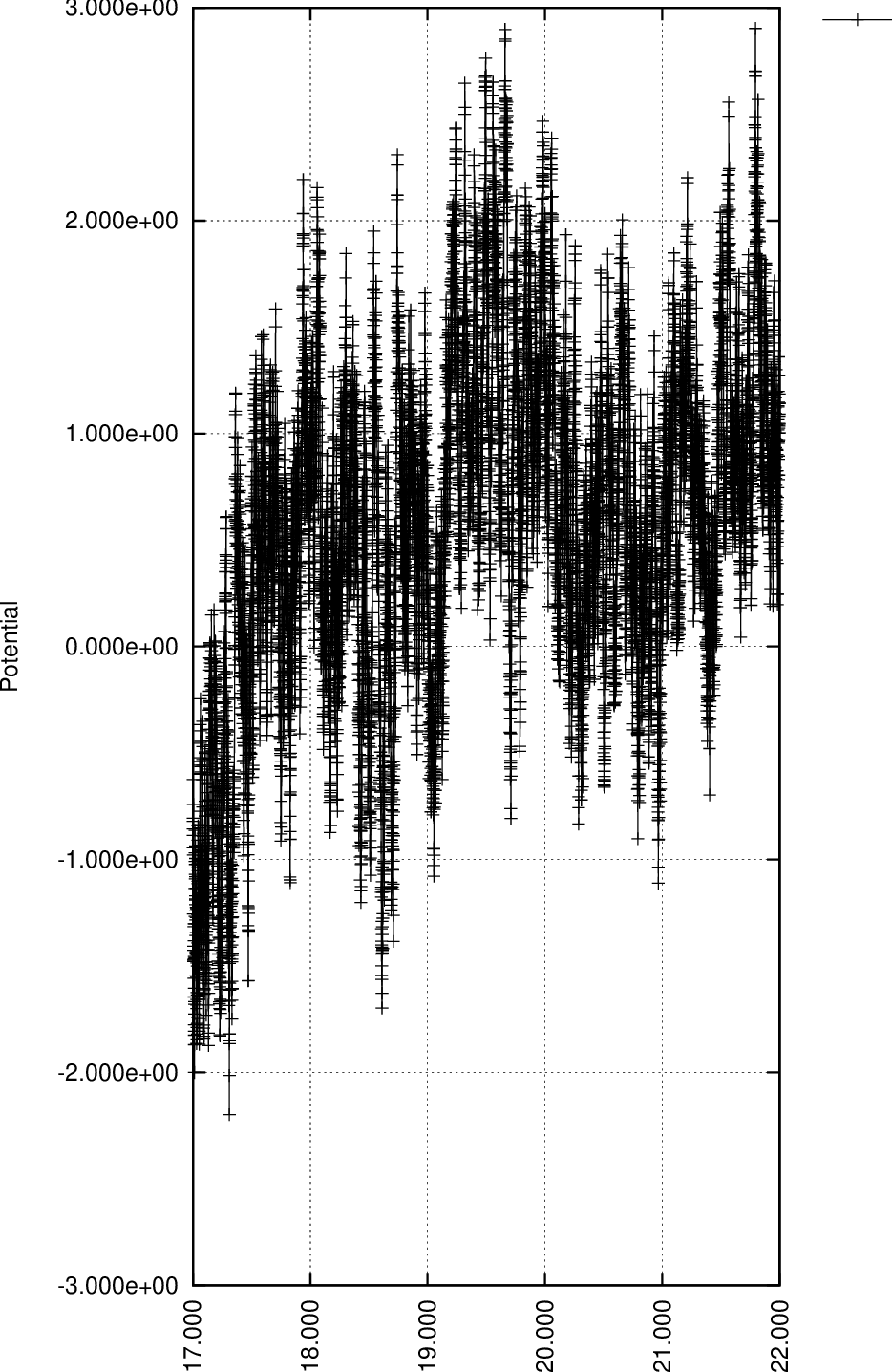}
        \caption{electric potential}%
        \label{fig:f1}
    \end{subfigure}
    \hfill
    \begin{subfigure}[b]
        {0.4\textwidth} \includegraphics[width=\textwidth]{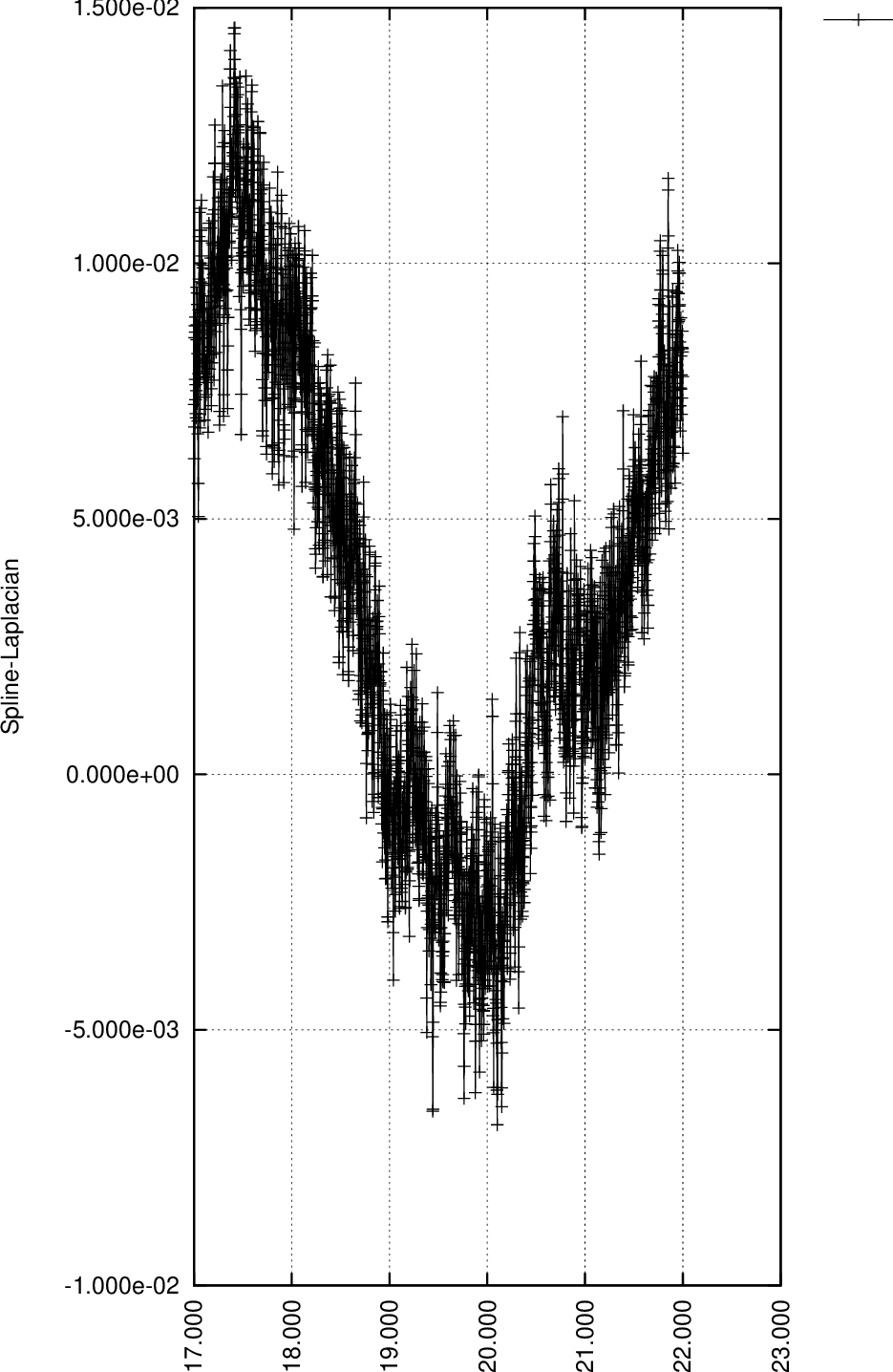}
        \caption{spline-Laplacian}%
        \label{fig:f2}
    \end{subfigure}
    \caption{Cumulative data over 20 subjects from EEG electrode site
    05-F3 is displayed:  (a) Raw electric potential data.  (b) Averaged
    spline-Laplacian data.}
\end{figure}
\subsection{All graphs}
A supplemental file contains graphs relevant to these calculations,
\begin{quote}
    https://www.ingber.com/smni16\_large-scale\_molecular\_EEGgraphs.pdf
\end{quote}

The first three pages are a pdf rendering of the 64-channel page in
\begin{quote}
    http://www.biosemi.com/download/Cap\_coords\_all.xls
\end{quote}

The pdf file contains are two sets of graphs, side by side, the raw
potentials and the spline-Laplacian transformed potentials:
\subsubsection{Raw Potential}
This set of 64 graphs contains all 245 runs from all 12 subjects of raw
electric potential data for each electrode site, within the time ranges
discussed in this paper.
\subsubsection{Spline-Laplacian}
This set of 64 graphs contains all 245 runs from all 12 subjects of
spline-Laplacian moving-averaged transformed data for each electrode
site, within the time ranges discussed in this paper.
\section{Fitting Theory To Data}
\subsection{Towards a zero-fit-parameter model}
In keeping more strictly with the SMNI zero-fit-parameter philosophy,
fits to data minimize the number of parameters to more strongly test
theory. Here, parameters are just the strength of regional connections.
Simple functional contributions of data (spline-Laplacian
transformations of scalp electric potential) dependence, drift-dependent
as discussed below, to $ B_{G^\prime{}}^{G} $ background were taken to
be 1/2 of the no-$ \mathbf{A} $ model. No additional parameters are
considered necessary to make the strong point that DSM is a better fit
to the data than CM, both of which are better than no CM. Therefore,
there also are no parameters scaling electric potentials to mesocolumnar
firings as in previous papers, but rather the data was used to determine
scales as discussed below. This was done in the context of using
spline-Laplacian transformations, to more strictly enforce the
correspondence between the transformed data and the mesocolumnar
firings, which may be electrode as well as subject dependent (due to
difference in skull/scalp properties, etc.).
\subsection{Refine data}
Instead of scaling columnar firings with parameters, for each subject,
for all runs using the chosen section of spline-Laplacian data, the
maximum and minimum values of data were used to scale maximum and
minimum firing, $ \pm N^{E} $ within each region, i.e., in each region,
each spline-Laplacian data point $ sl $ was scaled to firing states by
the factor $ M_{sl} $,

\begin{equation*}
    sl>0:M_{sl}=\min [N^{E},sl*N^{E}/(\max [\max (sl),|\min (sl)|])]
\end{equation*}

\begin{equation}
    sl<0:M_{sl}=\max [-N^{E},sl*N^{E}/(\max [\max (sl),|\min (sl)|])]
\end{equation}
\subsection{$ \mathbf{A} $ contribution to synaptic background}
The waves depend on aggregates of their $ \mathbf{\Pi }=\mathbf{p}+q\mathbf
{A} $ dynamics. E.g., this can be modeled as a Taylor expansion in $ |\mathbf
{A}| $,

\begin{equation}
    B_{G^\prime{}}^{G}\rightarrow B_{G^\prime{}}^{G}+\mathbf{A}B^\prime{}_
    {G^\prime{}}^{G}\:,\:B_{E^\prime{}}^{\ddagger E}=B_{E^\prime{}}^{\ddagger
    E}+\mathbf{A}B^\prime{}_{E^\prime{}}^{\ddagger E}
\end{equation}
For the $ \mathbf{A} $ model, only for $ B^\prime{}_{G^\prime{}}^{E} $
was added with a factor of the SMNI drift. The contribution of the
product $ \mathbf{A}B^\prime{}_{G^\prime{}}^{G} $ is taken to be a drift
factor multiplied by a factor of $ B_{G^\prime{}}^{G} $. The drift for
each region is not calculated from any trend of the data, but from the
SMNI nonlinear drift $ g^{G} $, as a moving average over the last 3
epochs representing about 6 msec. The moving-averaged drift numerically
fell between -0.5 and 0.5.
\subsection{Summary of proportionalities}
In summary,

\begin{equation*}
    M^{G}\propto\mathbf{I}
\end{equation*}

\begin{equation*}
    \mathrm{EEG}\propto\mathbf{I}
\end{equation*}

\begin{equation*}
    \mathbf{A}\propto\mathbf{I}
\end{equation*}

\begin{equation*}
    B^\prime{}_{G^\prime{}}^{E}\propto[\mathrm{Ca}^{2+}]
\end{equation*}

\begin{equation}
    [\mathrm{Ca}^{2+}]\propto\mathbf{A}
\end{equation}
where the last proportionality applies to the influence of $ \mathbf{A} $
on free $ [\mathrm{Ca}^{2+}] $ at synapses via $ \mathbf{\Pi }=\mathbf{p}+q\mathbf
{A} $.
\subsection{ASA}
The author's Adaptive Simulated Annealing (ASA) code
\citep{Ingber1993} is used for all optimization to fit SMNI parameters to
EEG data.

The main ASA OPTIONS that were turned on to tune the optimization were:
\begin{quote}
    ASA\_FUZZY \hfil\\
    QUENCH\_PARAMETERS with OPTIONS$ \rightarrow $User\_Quench\_Param\_Scale
    [] = 1.75 \hfil\\
    QUENCH\_COST with OPTIONS$ \rightarrow $User\_Quench\_Cost\_Scale =
    2 \hfil\\
    ASA\_QUEUE with OPTIONS$ \rightarrow $Queue\_Size = 5 \hfil\\
    OPTIONS$ \rightarrow $Limit\_Generated = 3000000
\end{quote}

The QUENCH OPTIONS were set to get good convergence to minima, tested
with longer runs without the QUENCHing within 1,000,000 generated
states.

The simplex code, contained in the asa\_usr.c file of the ASA code, was
run after ASA, permitting up to 5000 additional generated states. No
better solutions were obtained within max-min bounds of parameters.

Runs were performed on The Extreme Science and Engineering Discovery
Environment platforms, as described at XSEDE.org. XSEDE clock time
running 24 nodes under MPI, i.e., independent runs for each of 12
subjects, each subject run with A model and no-A model, was 37 hrs on
Stampede (Texas Advanced Computing Center (TACC) Dell PowerEdge C8220
Cluster with Intel Xeon Phi coprocessors) = 888 CPU hrs per run, or 48
hrs on Gordon (San Diego Supercomputer Center (SDSC) Appro with 8-core,
2.6-GHz Intel Sandy Bridge processors) = 1152 CPU hrs per run.

\subsection{Compare Testing with Training}
For each of the 12 subjects, it was possible to find 10 Training runs
and 10 Testing runs. Comparison with switched runs for Training and
Testing reveals some subjects with modest outlier runs; i.e., if they
were consistent across runs, then their Training cost functions should
be less than their Testing cost functions. The Table below presents the
results. As can be seen, the $ \mathbf{A} $ model clearly outperformed
the no-$ \mathbf{A} $ model. Note that cost functions with an $ |\mathbf
{A}| $ model are much worse that either the $ \mathbf{A} $ model or the
no-$ \mathbf{A} $ model. Runs with different signs on the drift and on
the absolute value of the drift also gave much higher cost functions
that the $ \mathbf{A} $ model.

\begin{table}
    \caption{Column 1 is the subject number; the other columns are cost
    functions.  Columns 2 and 3 are {no-\textbf{A}} model's Training (TR0)
    and Testing (TE0).  Columns 4 and 5 are {\textbf{A}} model's
    Training (TR\textbf{A})and Testing (TE\textbf{A}).  Columns 4 and 5
    are {no-\textbf{A}} model's Training (TRs0) and Testing (TEs0), with
    switched runs.  Columns 6 and 7 are {\textbf{A}} model's Training (TRs\textbf
    {A})and Testing (TEs\textbf{A}), with switched runs.  Columns 8 and
    9 are {\textbar\textbf{A}\textbar} model's Training (TR\textbar\textbf
    {A}\textbar) and Testing (TE\textbar\textbf{A}\textbar).}%
    \label{tab:t1} \begin{tabular}{|l|D{.}{.}{-1}D{.}{.}{-1}|D{.}{.}{-1}D{.}{.}{-1}|D{.}{.}{-1}D{.}{.}{-1}|D{.}{.}{-1}D{.}{.}{-1}|D{.}{.}{-1}D{.}{.}{-1}|}
        \hline
        \multicolumn{1}{|c}{Sub} & \multicolumn{1}{|c}{TR0} & \multicolumn{1}{c}{TE0} & \multicolumn{1}{|c}{TR\textbf{A}} & \multicolumn{1}{c}{TE\textbf{A}} & \multicolumn{1}{|c}{TRs0} & \multicolumn{1}{c}{TEs0} & \multicolumn{1}{|c}{TRs\textbf{A}} & \multicolumn{1}{c}{TEs\textbf{A}} & \multicolumn{1}{|c}{TR\textbar\textbf{A}\textbar} & \multicolumn{1}{c|}{TE\textbar\textbf{A}\textbar} \\ 
        \hline
        s01    & 85.61 & 121.4 & 62.08 & 97.31 & 120.3 & 86.85 & 96.62 & 62.90 & 98.17 & 132.3 \\ 
        s02    & 70.66 & 51.27 & 52.90 & 36.56 & 51.02 & 70.81 & 36.45 & 53.05 & 96.92 & 79.31 \\ 
        s03    & 61.26 & 79.58 & 43.12 & 55.63 & 78.96 & 61.43 & 55.24 & 43.22 & 83.74 & 104.5 \\ 
        s04    & 52.25 & 64.12 & 34.46 & 46.58 & 63.49 & 53.06 & 45.98 & 35.00 & 70.77 & 83.36 \\ 
        s05    & 67.20 & 72.22 & 47.29 & 51.44 & 71.30 & 67.78 & 51.14 & 47.64 & 85.52 & 88.36 \\ 
        s06    & 84.46 & 69.44 & 64.89 & 45.17 & 68.88 & 84.58 & 44.72 & 64.98 & 96.90 & 82.72 \\ 
        s07    & 68.60 & 78.62 & 49.67 & 56.87 & 78.37 & 68.76 & 56.77 & 49.76 & 87.00 & 95.64 \\ 
        s08    & 47.09 & 44.16 & 34.23 & 34.89 & 43.46 & 47.89 & 34.76 & 34.57 & 72.11 & 68.90 \\ 
        s09    & 47.52 & 25.22 & 39.19 & 16.47 & 24.96 & 48.06 & 16.29 & 39.74 & 85.37 & 66.76 \\ 
        s10    & 53.10 & 33.33 & 40.35 & 22.51 & 33.10 & 53.81 & 22.39 & 40.75 & 75.93 & 66.75 \\ 
        s11    & 43.91 & 51.15 & 33.21 & 37.64 & 50.93 & 44.38 & 37.52 & 33.50 & 70.90 & 87.31 \\ 
        s12    & 45.71 & 45.20 & 30.99 & 31.58 & 44.85 & 46.07 & 31.45 & 31.15 & 65.15 & 70.59 \\ 
        \hline
    \end{tabular}
\end{table}
\section{Conclusion}
An SMNI model has been developed to calculate coupling of molecular
scales of $ \mathrm{Ca}^{2+} $ wave dynamics with $ \mathbf{A} $ fields
developed at macroscopic regional scales measured by coherent neuronal
firing activity measured by scalp EEG during STM tasks. This requires
crossing molecular, microscopic (synaptic and neuronal), mesoscopic (minicolumns
and macrocolumns), and macroscopic regional scales.

Considerations of both classical and quantum physics give predictions of
the influence of $ \mathbf{A} $ on the momenta of $ \mathrm{Ca}^{2+} $
waves during STM processing as measured by scalp EEG\@. Since the spatial
scales of $ \mathrm{Ca}^{2+} $ wave and macro-EEG are quite disparate,
an experiment would have to be able to correlate both scales in time
scales on the order of tens of milliseconds.

This study is robust against much theoretical modeling, as experimental
data is used wherever possible. The theoretical construct of the
canonical momentum $ \mathbf{\Pi }=\mathbf{p}+q\mathbf{A} $ is firmly
entrenched in classical and quantum mechanics.

Previous calculations appeared to suffer from severe outliers in the EEG
data used to fit the SMNI models. Therefore new data was sought and
after a selection was made, care was taken to select sections of data
appropriate to the SMNI model, e.g., during P300 task, moving averages
consistent with SMNI time scales, etc. New fits indeed show that the $
\mathbf{A} $ is a much better fit to the data than the no-$ \mathbf{A} $
model.

\section*{Acknowledgment} \hfil\\
I thank the National Science Foundation's Extreme Science and
Engineering Discovery Environment (XSEDE.org), for three supercomputer
grants since February 2013, ``Electroencephalographic field influence on
calcium momentum waves'', one under PHY130022 and two under
TG-MCB140110. I thank Ramesh Srinivasan for giving me part of his Matlab
code for spline-Laplacian scalp transformations. \hfil\\

\clearpage
\section*{}
\bibliographystyle{elsarticle-harv.bst}
\bibliography{lingber}

\begin{thebibliography}{56}
\expandafter\ifx\csname natexlab\endcsname\relax\def\natexlab#1{#1}\fi
\providecommand{\url}[1]{\texttt{#1}}
\providecommand{\href}[2]{#2}
\providecommand{\path}[1]{#1}
\providecommand{\DOIprefix}{doi:}
\providecommand{\ArXivprefix}{arXiv:}
\providecommand{\URLprefix}{URL: }
\providecommand{\Pubmedprefix}{pmid:}
\providecommand{\doi}[1]{\href{http://dx.doi.org/#1}{\path{#1}}}
\providecommand{\Pubmed}[1]{\href{pmid:#1}{\path{#1}}}
\providecommand{\bibinfo}[2]{#2}
\ifx\xfnm\relax \def\xfnm[#1]{\unskip,\space#1}\fi
\bibitem[{Agulhon et~al.(2008)Agulhon, Petravicz, McMullen, Sweger, Minton,
  Taves, Casper, Fiacco and
  McCarthy}]{Agulhon+Petravicz+McMullen+Sweger+Minton+Taves+Casper+Fiacco+McCarthy2008}
\bibinfo{author}{Agulhon, C.}, \bibinfo{author}{Petravicz, J.},
  \bibinfo{author}{McMullen, A.}, \bibinfo{author}{Sweger, E.},
  \bibinfo{author}{Minton, S.}, \bibinfo{author}{Taves, S.},
  \bibinfo{author}{Casper, K.}, \bibinfo{author}{Fiacco, T.},
  \bibinfo{author}{McCarthy, K.}, \bibinfo{year}{2008}.
\newblock \bibinfo{title}{What is the role of astrocyte calcium in
  neurophysiology?}
\newblock \bibinfo{journal}{Neuron} \bibinfo{volume}{59},
  \bibinfo{pages}{932--946}.
\bibitem[{Asher(2012)}]{Asher2012}
\bibinfo{author}{Asher, J.}, \bibinfo{year}{2012}.
\newblock \bibinfo{title}{{Brain's} code for visual working memory deciphered
  in monkeys {NIH}-funded study}.
\newblock \bibinfo{type}{Technical Report} \bibinfo{number}{NIH Press Release}.
  NIH. \bibinfo{address}{Bethesda, MD}.
\newblock
  \bibinfo{note}{{}\url{http://www.nimh.nih.gov/news/science-news/2012/in-sync-brain-waves-hold-memory-of-objects-just-seen.shtml}}.
\bibitem[{Atmanspacher(2015)}]{Atmanspacher2015}
\bibinfo{author}{Atmanspacher, H.}, \bibinfo{year}{2015}.
\newblock \bibinfo{title}{Quantum approaches to consciousness}, in:
  \bibinfo{editor}{Zalta, E.} (Ed.), \bibinfo{booktitle}{The Stanford
  Encyclopedia of Philosophy}. \bibinfo{publisher}{Stanford U},
  \bibinfo{address}{Palo Alto}, p.~\bibinfo{pages}{1}.
\newblock
  \bibinfo{note}{{}\url{http://plato.stanford.edu/archives/sum2015/entries/qt-consciousness}}.
\bibitem[{Beck(2008)}]{Beck2008}
\bibinfo{author}{Beck, F.}, \bibinfo{year}{2008}.
\newblock \bibinfo{title}{Synaptic quantum tunnelling in brain activity}.
\newblock \bibinfo{journal}{Neuroquantology} \bibinfo{volume}{6},
  \bibinfo{pages}{140--151}.
\newblock \bibinfo{note}{{}\url{http://dx.doi.org/10.14704/nq.2008.6.2.168}}.
\bibitem[{Cheng(1972)}]{Cheng1972}
\bibinfo{author}{Cheng, K.}, \bibinfo{year}{1972}.
\newblock \bibinfo{title}{Quantization of a general dynamical system by
  {Feynman's} path integration formulation}.
\newblock \bibinfo{journal}{Journal of Mathematical Physics}
  \bibinfo{volume}{13}, \bibinfo{pages}{1723--1726}.
\bibitem[{Citi et~al.(2010)Citi, Poli and Cinel}]{Citi+Poli+Cinel2010}
\bibinfo{author}{Citi, L.}, \bibinfo{author}{Poli, R.}, \bibinfo{author}{Cinel,
  C.}, \bibinfo{year}{2010}.
\newblock \bibinfo{title}{Documenting, modelling and exploiting {P300}
  amplitude changes due to variable target delays in {Donchin's} speller}.
\newblock \bibinfo{journal}{Journal of Neural Engineering} \bibinfo{volume}{7},
  \bibinfo{pages}{1--21}.
\newblock \bibinfo{note}{{}\url{http://dx.doi.org/10.1088/1741-2560/7/5/056006}}.
\bibitem[{Facchi et~al.(2004)Facchi, Lidar and
  Pascazio}]{Facchi+Lidar+Pascazio2004}
\bibinfo{author}{Facchi, P.}, \bibinfo{author}{Lidar, D.},
  \bibinfo{author}{Pascazio, S.}, \bibinfo{year}{2004}.
\newblock \bibinfo{title}{Unification of dynamical decoupling and the quantum
  zeno effect}.
\newblock \bibinfo{journal}{Physical Review A} \bibinfo{volume}{69},
  \bibinfo{pages}{1--6}.
\bibitem[{Facchi and Pascazio(2008)}]{Facchi+Pascazio2008}
\bibinfo{author}{Facchi, P.}, \bibinfo{author}{Pascazio, S.},
  \bibinfo{year}{2008}.
\newblock \bibinfo{title}{Quantum zeno dynamics: mathematical and physical
  aspects}.
\newblock \bibinfo{journal}{Journal of Physics A} \bibinfo{volume}{41},
  \bibinfo{pages}{1--45}.
\bibitem[{Fiacco and McCarthy(2004)}]{Fiacco+McCarthy2004}
\bibinfo{author}{Fiacco, T.}, \bibinfo{author}{McCarthy, K.},
  \bibinfo{year}{2004}.
\newblock \bibinfo{title}{Intracellular astrocyte calcium waves in situ
  increase the frequency of spontaneous {AMPA} receptor currents in {CA1}
  pyramidal neurons}.
\newblock \bibinfo{journal}{Journal of Neuroscience} \bibinfo{volume}{24},
  \bibinfo{pages}{722--732}.
\bibitem[{Giacosa and Pagliara(2014)}]{Giacosa+Pagliara2014}
\bibinfo{author}{Giacosa, G.}, \bibinfo{author}{Pagliara, G.},
  \bibinfo{year}{2014}.
\newblock \bibinfo{title}{Quantum zeno effect by general measurements}.
\newblock \bibinfo{journal}{Physical Review A} \bibinfo{volume}{052107},
  \bibinfo{pages}{1--5}.
\bibitem[{Goldberger et~al.(2000)Goldberger, Amaral, Glass, Hausdorff, Ivanov,
  Mark, Mietus, Moody, Peng and
  Stanley}]{Goldberger+Amaral+Glass+Hausdorff+Ivanov+Mark+Mietus+Moody+Peng+Stanley2000}
\bibinfo{author}{Goldberger, A.}, \bibinfo{author}{Amaral, L.},
  \bibinfo{author}{Glass, L.}, \bibinfo{author}{Hausdorff, J.},
  \bibinfo{author}{Ivanov, P.}, \bibinfo{author}{Mark, R.},
  \bibinfo{author}{Mietus, J.}, \bibinfo{author}{Moody, G.},
  \bibinfo{author}{Peng, C.K.}, \bibinfo{author}{Stanley, H.},
  \bibinfo{year}{2000}.
\newblock \bibinfo{title}{{PhysioBank,} {PhysioToolkit,} and {PhysioNet:}
  components of a new research resource for complex physiologic signals}.
\newblock \bibinfo{journal}{Circulation} \bibinfo{volume}{101},
  \bibinfo{pages}{e215--e220}.
\newblock
  \bibinfo{note}{{}\url{http://circ.ahajournals.org/cgi/content/full/101/23/e215}}.
\bibitem[{Ingber(1982)}]{Ingber1982}
\bibinfo{author}{Ingber, L.}, \bibinfo{year}{1982}.
\newblock \bibinfo{title}{Statistical mechanics of neocortical interactions. i.
  basic formulation}.
\newblock \bibinfo{journal}{Physica D} \bibinfo{volume}{5},
  \bibinfo{pages}{83--107}.
\newblock \bibinfo{note}{{}\url{http://www.ingber.com/smni82\_basic.pdf}}.
\bibitem[{Ingber(1983)}]{Ingber1983}
\bibinfo{author}{Ingber, L.}, \bibinfo{year}{1983}.
\newblock \bibinfo{title}{Statistical mechanics of neocortical interactions.
  dynamics of synaptic modification}.
\newblock \bibinfo{journal}{Physical Review A} \bibinfo{volume}{28},
  \bibinfo{pages}{395--416}.
\newblock \bibinfo{note}{{}\url{http://www.ingber.com/smni83\_dynamics.pdf}}.
\bibitem[{Ingber(1984)}]{Ingber1984}
\bibinfo{author}{Ingber, L.}, \bibinfo{year}{1984}.
\newblock \bibinfo{title}{Statistical mechanics of neocortical interactions.
  derivation of short-term-memory capacity}.
\newblock \bibinfo{journal}{Physical Review A} \bibinfo{volume}{29},
  \bibinfo{pages}{3346--3358}.
\newblock \bibinfo{note}{{}\url{http://www.ingber.com/smni84\_stm.pdf}}.
\bibitem[{Ingber(1985a)}]{Ingber1985a}
\bibinfo{author}{Ingber, L.}, \bibinfo{year}{1985a}.
\newblock \bibinfo{title}{Statistical mechanics of neocortical interactions:
  Stability and duration of the {7+}-2 rule of short-term-memory capacity}.
\newblock \bibinfo{journal}{Physical Review A} \bibinfo{volume}{31},
  \bibinfo{pages}{1183--1186}.
\newblock \bibinfo{note}{{}\url{http://www.ingber.com/smni85\_stm.pdf}}.
\bibitem[{Ingber(1985b)}]{Ingber1985b}
\bibinfo{author}{Ingber, L.}, \bibinfo{year}{1985b}.
\newblock \bibinfo{title}{Statistical mechanics of neocortical interactions.
  {EEG} dispersion relations}.
\newblock \bibinfo{journal}{IEEE Transactions in Biomedical Engineering}
  \bibinfo{volume}{32}, \bibinfo{pages}{91--94}.
\newblock \bibinfo{note}{{}\url{http://www.ingber.com/smni85\_eeg.pdf}}.
\bibitem[{Ingber(1993)}]{Ingber1993}
\bibinfo{author}{Ingber, L.}, \bibinfo{year}{1993}.
\newblock \bibinfo{title}{Adaptive Simulated Annealing {(ASA)}}.
\newblock \bibinfo{type}{Technical Report} \bibinfo{number}{Global optimization
  C-code}. Caltech Alumni Association. \bibinfo{address}{Pasadena, CA}.
\newblock \bibinfo{note}{{}\url{http://www.ingber.com/\#ASA-CODE}}.
\bibitem[{Ingber(1994)}]{Ingber1994}
\bibinfo{author}{Ingber, L.}, \bibinfo{year}{1994}.
\newblock \bibinfo{title}{Statistical mechanics of neocortical interactions:
  Path-integral evolution of short-term memory}.
\newblock \bibinfo{journal}{Physical Review E} \bibinfo{volume}{49},
  \bibinfo{pages}{4652--4664}.
\newblock \bibinfo{note}{{}\url{http://www.ingber.com/smni94\_stm.pdf}}.
\bibitem[{Ingber(1997a)}]{Ingber1997a}
\bibinfo{author}{Ingber, L.}, \bibinfo{year}{1997a}.
\newblock \bibinfo{title}{Statistical mechanics of neocortical interactions:
  Applications of canonical momenta indicators to electroencephalography}.
\newblock \bibinfo{journal}{Physical Review E} \bibinfo{volume}{55},
  \bibinfo{pages}{4578--4593}.
\newblock \bibinfo{note}{{}\url{http://www.ingber.com/smni97\_cmi.pdf}}.
\bibitem[{Ingber(1997b)}]{Ingber1997b}
\bibinfo{author}{Ingber, L.}, \bibinfo{year}{1997b}.
\newblock \bibinfo{title}{{EEG} Database}.
\newblock \bibinfo{publisher}{UCI Machine Learning Repository},
  \bibinfo{address}{Irvine, CA}.
\newblock
  \bibinfo{note}{{}\url{http://archive.ics.uci.edu/ml/datasets/EEG+Database}}.
\bibitem[{Ingber(1998)}]{Ingber1998}
\bibinfo{author}{Ingber, L.}, \bibinfo{year}{1998}.
\newblock \bibinfo{title}{Statistical mechanics of neocortical interactions:
  Training and testing canonical momenta indicators of {EEG}}.
\newblock \bibinfo{journal}{Mathematical Computer Modelling}
  \bibinfo{volume}{27}, \bibinfo{pages}{33--64}.
\newblock \bibinfo{note}{{}\url{http://www.ingber.com/smni98\_cmi\_test.pdf}}.
\bibitem[{Ingber(2007)}]{Ingber2007}
\bibinfo{author}{Ingber, L.}, \bibinfo{year}{2007}.
\newblock \bibinfo{title}{Ideas by statistical mechanics {(ISM)}}.
\newblock \bibinfo{journal}{Journal Integrated Systems Design and Process
  Science} \bibinfo{volume}{11}, \bibinfo{pages}{31--54}.
\newblock \bibinfo{note}{{}Special Issue: Biologically Inspired Computing.}
\bibitem[{Ingber(2009)}]{Ingber2009}
\bibinfo{author}{Ingber, L.}, \bibinfo{year}{2009}.
\newblock \bibinfo{title}{Statistical mechanics of neocortical interactions:
  Nonlinear columnar electroencephalography}.
\newblock \bibinfo{journal}{NeuroQuantology Journal} \bibinfo{volume}{7},
  \bibinfo{pages}{500--529}.
\newblock
  \bibinfo{note}{{}\url{http://www.ingber.com/smni09\_nonlin\_column\_eeg.pdf}}.
\bibitem[{Ingber(2011)}]{Ingber2011}
\bibinfo{author}{Ingber, L.}, \bibinfo{year}{2011}.
\newblock \bibinfo{title}{Computational algorithms derived from multiple scales
  of neocortical processing}, in: \bibinfo{editor}{Pereira,~Jr., A.},
  \bibinfo{editor}{Massad, E.}, \bibinfo{editor}{Bobbitt, N.} (Eds.),
  \bibinfo{booktitle}{Pointing at Boundaries: Integrating Computation and
  Cognition on Biological Grounds}. \bibinfo{publisher}{Springer},
  \bibinfo{address}{New York}, pp. \bibinfo{pages}{1--13}.
\newblock \bibinfo{note}{{}Invited Paper.
  \url{http://www.ingber.com/smni11\_cog\_comp.pdf and
  http://dx.doi.org/10.1007/s12559-011-9105-4}}.
\bibitem[{Ingber(2012)}]{Ingber2012}
\bibinfo{author}{Ingber, L.}, \bibinfo{year}{2012}.
\newblock \bibinfo{title}{Columnar {EEG} magnetic influences on molecular
  development of short-term memory}, in: \bibinfo{editor}{Kalivas, G.},
  \bibinfo{editor}{Petralia, S.} (Eds.), \bibinfo{booktitle}{Short-Term Memory:
  New Research}. \bibinfo{publisher}{Nova}, \bibinfo{address}{Hauppauge, NY},
  pp. \bibinfo{pages}{37--72}.
\newblock \bibinfo{note}{{}Invited Paper.
  \url{http://www.ingber.com/smni11\_stm\_scales.pdf}}.
\bibitem[{Ingber(2015)}]{Ingber2015}
\bibinfo{author}{Ingber, L.}, \bibinfo{year}{2015}.
\newblock \bibinfo{title}{Calculating consciousness correlates at multiple
  scales of neocortical interactions}, in: \bibinfo{editor}{Costa, A.},
  \bibinfo{editor}{Villalba, E.} (Eds.), \bibinfo{booktitle}{Horizons in
  Neuroscience Research}. \bibinfo{publisher}{Nova},
  \bibinfo{address}{Hauppauge, NY}, pp. \bibinfo{pages}{153--186}.
\newblock \bibinfo{note}{{}ISBN: 978-1-63482-632-7. Invited paper.
  \url{http://www.ingber.com/smni15\_calc\_conscious.pdf}}.
\bibitem[{Ingber et~al.(2001)Ingber, Chen, Mondescu, Muzzall and
  Renedo}]{Ingber+Chen+Mondescu+Muzzall+Renedo2001}
\bibinfo{author}{Ingber, L.}, \bibinfo{author}{Chen, C.},
  \bibinfo{author}{Mondescu, R.}, \bibinfo{author}{Muzzall, D.},
  \bibinfo{author}{Renedo, M.}, \bibinfo{year}{2001}.
\newblock \bibinfo{title}{Probability tree algorithm for general diffusion
  processes}.
\newblock \bibinfo{journal}{Physical Review E} \bibinfo{volume}{64},
  \bibinfo{pages}{056702--056707}.
\newblock \bibinfo{note}{{}\url{http://www.ingber.com/path01\_pathtree.pdf}}.
\bibitem[{Ingber and Nunez(1990)}]{Ingber+Nunez1990}
\bibinfo{author}{Ingber, L.}, \bibinfo{author}{Nunez, P.},
  \bibinfo{year}{1990}.
\newblock \bibinfo{title}{Multiple scales of statistical physics of neocortex:
  Application to electroencephalography}.
\newblock \bibinfo{journal}{Mathematical Computer Modelling}
  \bibinfo{volume}{13}, \bibinfo{pages}{83--95}.
\bibitem[{Ingber and Nunez(1995)}]{Ingber+Nunez1995}
\bibinfo{author}{Ingber, L.}, \bibinfo{author}{Nunez, P.},
  \bibinfo{year}{1995}.
\newblock \bibinfo{title}{Statistical mechanics of neocortical interactions:
  High resolution path-integral calculation of short-term memory}.
\newblock \bibinfo{journal}{Physical Review E} \bibinfo{volume}{51},
  \bibinfo{pages}{5074--5083}.
\newblock \bibinfo{note}{{}\url{http://www.ingber.com/smni95\_stm.pdf}}.
\bibitem[{Ingber and Nunez(2010)}]{Ingber+Nunez2010}
\bibinfo{author}{Ingber, L.}, \bibinfo{author}{Nunez, P.},
  \bibinfo{year}{2010}.
\newblock \bibinfo{title}{Neocortical dynamics at multiple scales: {EEG}
  standing waves, statistical mechanics, and physical analogs}.
\newblock \bibinfo{journal}{Mathematical Biosciences} \bibinfo{volume}{229},
  \bibinfo{pages}{160--173}.
\newblock
  \bibinfo{note}{{}\url{http://www.ingber.com/smni10\_multiple\_scales.pdf}}.
\bibitem[{Ingber et~al.(2014)Ingber, Pappalepore and
  Stesiak}]{Ingber+Pappalepore+Stesiak2014}
\bibinfo{author}{Ingber, L.}, \bibinfo{author}{Pappalepore, M.},
  \bibinfo{author}{Stesiak, R.}, \bibinfo{year}{2014}.
\newblock \bibinfo{title}{Electroencephalographic field influence on calcium
  momentum waves}.
\newblock \bibinfo{journal}{Journal of Theoretical Biology}
  \bibinfo{volume}{343}, \bibinfo{pages}{138--153}.
\newblock \bibinfo{note}{{}\url{http://www.ingber.com/smni14\_eeg\_ca.pdf and
  http://dx.doi.org/10.1016/j.jtbi.2013.11.002}}.
\bibitem[{Innocenti et~al.(2000)Innocenti, Parpura and
  Haydon}]{Innocenti+Parpura+Haydon2000}
\bibinfo{author}{Innocenti, B.}, \bibinfo{author}{Parpura, V.},
  \bibinfo{author}{Haydon, P.}, \bibinfo{year}{2000}.
\newblock \bibinfo{title}{Imaging extracellular waves of glutamate during
  calcium signaling in cultured astrocytes}.
\newblock \bibinfo{journal}{Journal of Neuroscience} \bibinfo{volume}{20},
  \bibinfo{pages}{1800--1808}.
\bibitem[{Kozlowski et~al.(2015)Kozlowski, Caballero-Benitez and
  Mekhov}]{Kozlowski+Caballero-Benitez+Mekhov2015}
\bibinfo{author}{Kozlowski, W.}, \bibinfo{author}{Caballero-Benitez, S.},
  \bibinfo{author}{Mekhov, I.}, \bibinfo{year}{2015}.
\newblock \bibinfo{title}{Non-hermitian dynamics in the quantum Zeno limit}.
\newblock \bibinfo{type}{Technical Report} \bibinfo{number}{arXiv:1510.04857
  [quant-ph]}. U Oxford. \bibinfo{address}{Oxford, UK}.
\bibitem[{Laloux et~al.(1999)Laloux, Cizeau, Bouchaud and
  Potters}]{Laloux+Cizeau+Bouchaud+Potters1999}
\bibinfo{author}{Laloux, L.}, \bibinfo{author}{Cizeau, P.},
  \bibinfo{author}{Bouchaud, J.}, \bibinfo{author}{Potters, M.},
  \bibinfo{year}{1999}.
\newblock \bibinfo{title}{Noise dressing of financial correlation matrices}.
\newblock \bibinfo{journal}{Physical Review Letters} \bibinfo{volume}{83},
  \bibinfo{pages}{1467--1470}.
\bibitem[{Langouche et~al.(1982)Langouche, Roekaerts and
  Tirapegui}]{Langouche+Roekaerts+Tirapegui1982}
\bibinfo{author}{Langouche, F.}, \bibinfo{author}{Roekaerts, D.},
  \bibinfo{author}{Tirapegui, E.}, \bibinfo{year}{1982}.
\newblock \bibinfo{title}{Functional Integration and Semiclassical Expansions}.
\newblock \bibinfo{publisher}{Reidel}, \bibinfo{address}{Dordrecht, The
  Netherlands}.
\bibitem[{Litterman and Winkelmann(1998)}]{Litterman+Winkelmann1998}
\bibinfo{author}{Litterman, R.}, \bibinfo{author}{Winkelmann, K.},
  \bibinfo{year}{1998}.
\newblock \bibinfo{title}{Estimating covariance matrices}.
\newblock \bibinfo{type}{Technical Report} \bibinfo{number}{Report}. Goldman
  Sachs. \bibinfo{address}{New York}.
\bibitem[{Lynch(1992)}]{Lynch1992}
\bibinfo{author}{Lynch, R.}, \bibinfo{year}{1992}.
\newblock \bibinfo{title}{Fundamental solutions of nine-point discrete
  laplacians}.
\newblock \bibinfo{journal}{Applied Numerical Mathematics}
  \bibinfo{volume}{10}, \bibinfo{pages}{325--334}.
\bibitem[{McKemmish et~al.(2009)McKemmish, Reimers, McKenzie, Mark and
  Hush}]{McKemmish+Reimers+McKenzie+Mark+Hush2009}
\bibinfo{author}{McKemmish, L.}, \bibinfo{author}{Reimers, J.},
  \bibinfo{author}{McKenzie, R.}, \bibinfo{author}{Mark, A.},
  \bibinfo{author}{Hush, N.}, \bibinfo{year}{2009}.
\newblock \bibinfo{title}{Penrose-hameroff orchestrated objective-reduction
  proposal for human consciousness is not biologically feasible}.
\newblock \bibinfo{journal}{Physical Review E} \bibinfo{volume}{80},
  \bibinfo{pages}{1--6}.
\newblock
  \bibinfo{note}{{}\url{http://link.aps.org/doi/10.1103/PhysRevE.80.021912}}.
\bibitem[{Mitterauer and
  Kofler-Westergren(2011)}]{Mitterauer+Kofler-Westergren2011}
\bibinfo{author}{Mitterauer, B.}, \bibinfo{author}{Kofler-Westergren, B.},
  \bibinfo{year}{2011}.
\newblock \bibinfo{title}{Possible effects of synaptic imbalances on
  oligodendrocyte-axonic interactions in schizophrenia: A hypothetical model}.
\newblock \bibinfo{journal}{Frontiers in Psychiatry} \bibinfo{volume}{2},
  \bibinfo{pages}{1--13}.
\newblock \bibinfo{note}{{}\url{http://doi.org/10.3389/fpsyt.2011.00015}}.
\bibitem[{Murakami and Okada(2006)}]{Murakami+Okada2006}
\bibinfo{author}{Murakami, S.}, \bibinfo{author}{Okada, Y.},
  \bibinfo{year}{2006}.
\newblock \bibinfo{title}{Contributions of principal neocortical neurons to
  magnetoencephalography and electroencephalography signals}.
\newblock \bibinfo{journal}{Journal of Physiology} \bibinfo{volume}{575},
  \bibinfo{pages}{925--936}.
\bibitem[{Nunez et~al.(2013)Nunez, Srinivasan and
  Ingber}]{Nunez+Srinivasan+Ingber2013}
\bibinfo{author}{Nunez, P.}, \bibinfo{author}{Srinivasan, R.},
  \bibinfo{author}{Ingber, L.}, \bibinfo{year}{2013}.
\newblock \bibinfo{title}{Theoretical and experimental electrophysiology in
  human neocortex: Multiscale correlates of conscious experience}, in:
  \bibinfo{editor}{Pesenson, M.} (Ed.), \bibinfo{booktitle}{Multiscale Analysis
  and Nonlinear Dynamics: From genes to the brain}. \bibinfo{publisher}{Wiley},
  \bibinfo{address}{New York}, pp. \bibinfo{pages}{149--178}.
\newblock \bibinfo{note}{{}\url{http://dx.doi.org/10.1002/9783527671632.ch06}}.
\bibitem[{Nunez and Srinivasan(2006)}]{Nunez+Srinivasan2006}
\bibinfo{author}{Nunez, P.L.}, \bibinfo{author}{Srinivasan, R.},
  \bibinfo{year}{2006}.
\newblock \bibinfo{title}{Electric Fields of the Brain: The Neurophysics of
  {EEG,} 2nd Ed}.
\newblock \bibinfo{publisher}{Oxford University Press},
  \bibinfo{address}{London}.
\bibitem[{Patil et~al.(2015)Patil, Chakram and
  Vengalattore}]{Patil+Chakram+Vengalattore2015}
\bibinfo{author}{Patil, Y.}, \bibinfo{author}{Chakram, S.},
  \bibinfo{author}{Vengalattore, M.}, \bibinfo{year}{2015}.
\newblock \bibinfo{title}{Measurement-induced localization of an ultracold
  lattice gas}.
\newblock \bibinfo{journal}{Physical Review Letters} \bibinfo{volume}{115},
  \bibinfo{pages}{1--5}.
\newblock
  \bibinfo{note}{{}\url{http://link.aps.org/doi/10.1103/PhysRevLett.115.140402}}.
\bibitem[{Pereira and Furlan(2009)}]{Pereira+Furlan2009}
\bibinfo{author}{Pereira,~Jr., A.}, \bibinfo{author}{Furlan, F.A.},
  \bibinfo{year}{2009}.
\newblock \bibinfo{title}{On the role of synchrony for neuron-astrocyte
  interactions and perceptual conscious processing}.
\newblock \bibinfo{journal}{Journal of Biological Physics}
  \bibinfo{volume}{35}, \bibinfo{pages}{465--480}.
\bibitem[{Ross(2012)}]{Ross2012}
\bibinfo{author}{Ross, W.}, \bibinfo{year}{2012}.
\newblock \bibinfo{title}{Understanding calcium waves and sparks in central
  neurons}.
\newblock \bibinfo{journal}{Nature Reviews Neuroscience} \bibinfo{volume}{13},
  \bibinfo{pages}{157--168}.
\bibitem[{Salazar et~al.(2012)Salazar, Dotson, Bressler and
  Gray}]{Salazar+Dotson+Bressler+Gray2012}
\bibinfo{author}{Salazar, R.}, \bibinfo{author}{Dotson, N.},
  \bibinfo{author}{Bressler, S.}, \bibinfo{author}{Gray, C.},
  \bibinfo{year}{2012}.
\newblock \bibinfo{title}{Content-specific fronto-parietal synchronization
  during visual working memory}.
\newblock \bibinfo{journal}{Science} \bibinfo{volume}{338},
  \bibinfo{pages}{1097--1100}.
\newblock \bibinfo{note}{{}\url{http://dx.doi.org/10.1126/science.1224000}}.
\bibitem[{Scemes and Giaume(2006)}]{Scemes+Giaume2006}
\bibinfo{author}{Scemes, E.}, \bibinfo{author}{Giaume, C.},
  \bibinfo{year}{2006}.
\newblock \bibinfo{title}{Astrocyte calcium waves: What they are and what they
  do}.
\newblock \bibinfo{journal}{Glia} \bibinfo{volume}{54},
  \bibinfo{pages}{716--725}.
\newblock \bibinfo{note}{{}\url{http://dx.doi.org/10.1002/glia.20374}}.
\bibitem[{Shaw and Vasudevan(1974)}]{Shaw+Vasudevan1974}
\bibinfo{author}{Shaw, G.}, \bibinfo{author}{Vasudevan, R.},
  \bibinfo{year}{1974}.
\newblock \bibinfo{title}{Persistent states of neural networks and the random
  nature of synaptic transmission}.
\newblock \bibinfo{journal}{Mathematical Biosciences} \bibinfo{volume}{21},
  \bibinfo{pages}{207--218}.
\bibitem[{Srinivasan et~al.(2006)Srinivasan, Winter and
  Nunez}]{Srinivasan+Winter+Nunez2006}
\bibinfo{author}{Srinivasan, R.}, \bibinfo{author}{Winter, W.},
  \bibinfo{author}{Nunez, P.}, \bibinfo{year}{2006}.
\newblock \bibinfo{title}{Source analysis of {EEG} oscillations using
  high-resolution {EEG} and {MEG}}.
\newblock \bibinfo{journal}{Progress Brain Research} \bibinfo{volume}{159},
  \bibinfo{pages}{29--42}.
\bibitem[{Volterra et~al.(2014)Volterra, Liaudet and
  Savtchouk}]{Volterra+Liaudet+Savtchouk2014}
\bibinfo{author}{Volterra, A.}, \bibinfo{author}{Liaudet, N.},
  \bibinfo{author}{Savtchouk, I.}, \bibinfo{year}{2014}.
\newblock \bibinfo{title}{Astrocyte {Ca2+} signalling: an unexpected
  complexity}.
\newblock \bibinfo{journal}{Nature Reviews Neuroscience} \bibinfo{volume}{15},
  \bibinfo{pages}{327--335}.
\bibitem[{Wu et~al.(2012)Wu, Wang and Yi}]{Wu+Wang+Yi2012}
\bibinfo{author}{Wu, S.}, \bibinfo{author}{Wang, L.}, \bibinfo{author}{Yi, X.},
  \bibinfo{year}{2012}.
\newblock \bibinfo{title}{Time-dependent decoherence-free subspace}.
\newblock \bibinfo{journal}{Journal of Physics A} \bibinfo{volume}{405305},
  \bibinfo{pages}{1--11}.
\bibitem[{Zhang et~al.(2014)Zhang, Ai, Li, Xu and Sun}]{Zhang+Ai+Li+Xu+Sun2014}
\bibinfo{author}{Zhang, P.}, \bibinfo{author}{Ai, Q.}, \bibinfo{author}{Li,
  Y.}, \bibinfo{author}{Xu, D.}, \bibinfo{author}{Sun, C.},
  \bibinfo{year}{2014}.
\newblock \bibinfo{title}{Dynamics of quantum zeno and anti-zeno effects in an
  open system}.
\newblock \bibinfo{journal}{Science China Physics, Mechanics Astronomy}
  \bibinfo{volume}{57}, \bibinfo{pages}{194--207}.
\newblock \bibinfo{note}{{}\url{http://dx.doi.org/10.1007/s11433-013-5377-x}}.
\bibitem[{Zhang et~al.(1997a)Zhang, Begleiter and
  Porjesz}]{Zhang+Begleiter+Porjesz1997}
\bibinfo{author}{Zhang, X.}, \bibinfo{author}{Begleiter, H.},
  \bibinfo{author}{Porjesz, B.}, \bibinfo{year}{1997}a.
\newblock \bibinfo{title}{Do chronic alcoholics have intact implicit memory? an
  {ERP} study}.
\newblock \bibinfo{journal}{Electroencephalography Clinical Neurophysiology}
  \bibinfo{volume}{103}, \bibinfo{pages}{457--473}.
\bibitem[{Zhang et~al.(1997b)Zhang, Begleiter, Porjesz and
  Litke}]{Zhang+Begleiter+Porjesz+Litke1997}
\bibinfo{author}{Zhang, X.}, \bibinfo{author}{Begleiter, H.},
  \bibinfo{author}{Porjesz, B.}, \bibinfo{author}{Litke, A.},
  \bibinfo{year}{1997}b.
\newblock \bibinfo{title}{Electrophysiological evidence of memory impairment in
  alcoholic patients}.
\newblock \bibinfo{journal}{Biological Psychiatry} \bibinfo{volume}{42},
  \bibinfo{pages}{1157--1171}.
\bibitem[{Zhang et~al.(1995)Zhang, Begleiter, Porjesz, Wang and
  Litke}]{Zhang+Begleiter+Porjesz+Wang+Litke1995}
\bibinfo{author}{Zhang, X.}, \bibinfo{author}{Begleiter, H.},
  \bibinfo{author}{Porjesz, B.}, \bibinfo{author}{Wang, W.},
  \bibinfo{author}{Litke, A.}, \bibinfo{year}{1995}.
\newblock \bibinfo{title}{Event related potentials during object recognition
  tasks}.
\newblock \bibinfo{journal}{Brain Research Bulletin} \bibinfo{volume}{38},
  \bibinfo{pages}{531--538}.
\bibitem[{Zhou et~al.(2015)Zhou, Mowrey, Tang and Xu}]{Zhou+Mowrey+Tang+Xu2015}
\bibinfo{author}{Zhou, D.}, \bibinfo{author}{Mowrey, D.},
  \bibinfo{author}{Tang, P.}, \bibinfo{author}{Xu, Y.}, \bibinfo{year}{2015}.
\newblock \bibinfo{title}{Percolation model of sensory transmission and loss of
  consciousness under general anesthesia}.
\newblock \bibinfo{journal}{Physical Review Letters} \bibinfo{volume}{115},
  \bibinfo{pages}{1--6}.
\newblock
  \bibinfo{note}{{}\url{http://link.aps.org/doi/10.1103/PhysRevLett.115.108103}}.

\end{thebibliography}
\ %
\end{document}